# Generation of In-group Asset Condition Data for Power System Reliability Assessment

Ming Dong, *Senior Member*, *IEEE*, Alexandre B. Nassif, *Senior Member*, *IEEE*, Wenyuan Li,
*Life Fellow*, *IEEE*,

*Abstract*— In a power system, unlike some critical and standalone assets that are equipped with condition monitoring devices, the conditions of most regular in-group assets are acquired through periodic inspection work. Due to their large quantities, significant amount of manual inspection effort and sometimes data management problems, it is not uncommon to find that the asset condition data in a target study area is unavailable or incomplete. Lack of asset condition data undermines the reliability assessment work. To solve this data problem and enhance data availability, this paper explores an unconventional method – generating numerical and non-numerical asset condition data based on condition degradation, condition correlation and categorical distribution models. Empirical knowledge from human experts can also be incorporated in the modeling process. Also, a probabilistic diversification step can be taken to make the generated numerical condition data probabilistic. This method can generate close-to-real asset condition data and has been validated systematically based on two public datasets. An area reliability assessment example based on cables is given to demonstrate the usefulness of this method with its generated data. This method can also be used to conveniently generate hypothetical asset condition data for research purposes.

## I. Introduction

POWER systems are asset intensive. In recent years, the regulatory requirements on many utilities in the world have increased. It is now expected to achieve a reasonable balance between the reliability of power supply and the cost of customers [1]. To quantitatively guide and manage such effort, accurate reliability assessment work that covers asset health prediction, system reliability prediction and asset management optimization has become very important. For such assessment work, the condition data of power assets is critical. Generally, based on the acquisition and availability of asset condition data, power assets can be classified into the following two kinds [2]:

1. Standalone assets equipped with condition monitoring devices: these assets are often critical assets in a power system. They can be costly assets such as some power generators, power transformers, HVDC converter stations and circuit breakers. They can also be special cables and overhead lines that serve critical locations of a system. For such assets, on the one hand, their importance can justify the installation of condition monitoring devices which are used to monitor, analyze and track their reliability statuses; on the other hand, they do not form up large asset groups due to their uniqueness in function, environment or manufacturing as well as their scarcity.

2. In-group assets that rely on periodic inspection work: these assets are the vast majority of assets in a power system. Typical examples are ordinary underground cables, conductors, poles, steel towers, service transformers, switchgears and so on. In reliability studies, many assets can be grouped together – the same type of assets that serve in the same or similar areas can often be considered as having the same electrical and mechanical condition degradation characteristics. Since it is often cost prohibitive to install condition monitoring devices on such assets due to their large quantities, utility companies rely on periodic inspection work to obtain their condition data at an interval of typically a few years.

Unfortunately, unlike standalone devices, the asset condition data for in-group assets is often unavailable or incomplete for a target study area. The following reasons could cause such a data problem: due to the large quantity of assets in the system, the periodic inspection work slowly rotates through the system. For example, in many North American utility companies, wood poles are inspected every 10 years [3] and cables are inspected every 5 years. As a result, it is very possible that the inspection for the target study area was done a few years ago and there is no updated condition data that reflect the current or future asset health statuses; since inspection work often requires a significant amount of manual effort, the inspection of assets in some areas are not conducted proactively. An asset only gets inspected if a certain problem is discovered and reported; in many places, computerized asset management systems were installed and applied not a long time ago. This means that some inspection work may identify and fix asset issues right away but the inspection records may not be properly kept for future study purposes.

Because of such common data availability issues, for in-group asset reliability studies, utility companies often have to rely on only asset age information which is readily available and can be read off from the asset labels if needed. Age based Weibull distribution model has been widely used for in-group asset reliability studies for a long time. The model only uses

M. Dong is with Department of Grid Reliability, Alberta Electric System Operator (AESO), Calgary, AB, Canada, T2P 0L4 (corresponding e-mail: mingdong@ieee.org); A.B. Nassif is with Asset Management, LUMA Energy, PR, United States, 00907; W.Li is with School of Electrical Engineering, Chongqing University, Chongqing,400044,China.

the asset age as input and assumes that all assets at the same age in an asset group have the same aging failure probability characterized by the Weibull degradation curve [5-7]. As discussed and shown in [2], this method ignores the differences of individual assets at the same age and can overly simplify the aging process. In some cases, the results can significantly deviate from the reality. To counter this problem, [2,4] proposed the use of functional age or conditional age to incorporate both asset age data and condition data and the failure probability assessment accuracy can be greatly enhanced. However, such methods cannot be effectively applied if the in-group asset condition data is unavailable or incomplete.

To address such a data problem that prevents more accurate asset reliability assessment work, this paper proposes a novel and unconventional method to enhance the availability of in-group asset condition data. The method can generate asset condition data based on condition degradation, condition correlation and categorical distribution models. Empirical knowledge from human experts can also be incorporated in the models. In addition, a probabilistic diversification step is used to make the generated data probabilistic. The significances of the proposed method are:

1. Based on limited existing data and/or empirical expert knowledge, current and future in-group asset conditions can be generated. This can effectively support the asset reliability assessment work such as asset and system reliability risk quantification and asset management optimization;
2. It can also help researchers who want to conduct research on topics related to reliability studies and asset management optimization. Currently, the public data in the field is limited and by using the proposed method, hypothetical asset condition data can be generated conveniently.

This paper is organized as follows: in Section II, the topic of modelling numerical asset conditions is discussed systematically. Four typical condition degradation models are reviewed and a condition correlation model is proposed to model the relationship between a target condition and its relevant conditions. Then the method to combine the condition degradation and condition correlation models as well as the process of determining the model parameters are explained. Furthermore, a probabilistic diversification step is proposed to make the generated numerical data probabilistic; in Section III, the method of using categorical distribution to probabilistically model non-numerical asset conditions is discussed; in Section IV, for validation purposes, data generated by the proposed method and compared to real data based on two public datasets from a few different perspectives; in Section V, an actual application example of area reliability assessment is given to demonstrate the usefulness of this approach and the generated data.

## II. ASSET NUMERICAL CONDITION MODELS

In this section, we discuss a few condition degradation models that assume the condition value change is driven by the asset service age. During inspection, one type of asset can have a few different condition attributes $C_1, C_2, C_3$ ... to describe the asset health from different perspectives. It should be noted that the conditions captured by regular inspections are non-destructive conditions. It should be noted that this paper does not consider destructive conditions because these conditions are obtained through destructive tests on limited samples and such samples will terminate their service after getting destructive tests [8]. Examples of such conditions are the structural/mechanical strength of towers and conductors, the insulation breakdown voltage and etc.; non-destructive conditions can exist in the form of either numerical or non-numerical values [2]. Non-numerical conditions often come from asset inspectors' judgement and are in the form of discrete ratings such as low to high and are going to be discussed in Section III.

### A. Numerical Condition Degradation Models

As [8-14] discussed, for numerical non-destructive conditions that are obtained from regular inspections, four commonly used statistical models can be applied to describe the degradation processes, i.e. the linear model, the exponential model, the logarithmic model and the power model, each of which is discussed as follows.

- Linear Model:

The linear model can be used to describe a steady degradation process. Some numerical conditions change at a steady rate with time. This is often the case for unprotected assets that are exposed to external risks. For example, for wood power poles, animal damages such as wood pecker holes can gradually accumulate with time; weather caused wood pole surface damage can deteriorate with outdoor exposure time [15]. The linear model is presented by the equation below:

$$C(t) = a \times t + b, \ t \geq 0 \quad (1)$$

Where $t$ is the service age, and $a$ and $b$ are two factors. The linear model in Fig. 1 shows a linear degradation process ($a = 2$ and $b = 3$).

- Exponential Model:

Unlike the linear model with a steady degradation rate, accelerated degradation can happen to some power asset conditions at their late life stage. For example, after the initial protective coating disappears, the corrosion process of steel products can be significantly accelerated [16]; the decreasing capacitance of capacitors over time is another example [8]. In such cases, the exponential model can be applied. It is presented by the equation below:

$$C(t) = b \times e^{at}, t \geq 0 \quad (2)$$

The exponential model in Fig. 1 shows an exponential degradation process ($a = 0.08$ and $b = 0.1$).

- Logarithmic Model

In contrast with the exponential model, the logarithmic model is applied for asset conditions with a faster degradation process at their early life stage. Resistance and transistor gains with time are two examples [8]. The logarithmic model is presented by the equation below:

$$C(t) = a \times ln(t) + b, t \geq 0 \quad (3)$$

The logarithmic model in Fig. 1 shows an logarithmic degradation process ($a = 25$ and $b = 1.5$).

- Power Model

With different sets of parameters, the power model can reveal accelerated degradation at either an early life stage or a late life stage. It is presented by the equation below:

$$C(t) = b \times t^a, t \geq 0 \quad (4)$$

when $a > 1$, the process is similar to the exponential model with accelerated degradation at a late life stage; when $a < 1$, the process is similar to the logarithmic model with accelerated degradation at an early life stage; when $a = 1$, the power model is simplified to a linear model. Fig. 1 illustrates two power models with $a > 1$ and $a < 1$.

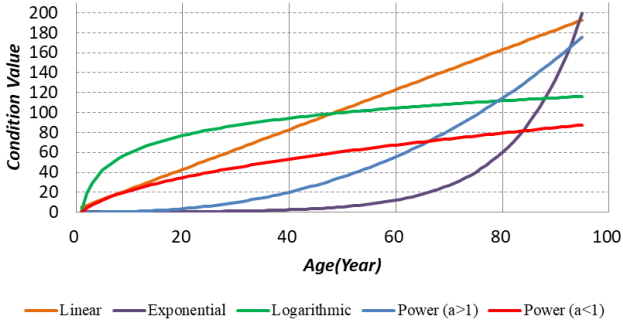

Fig.1. Numerical Condition Degradation Models

### B. Asset Condition Correlation Model

Different from the above age-driven models, the condition correlation model focuses on the relationship of the current condition values and the condition values obtained from the last asset inspection. This is applicable when there are some historical condition data for at least two continuous inspection years. In such cases, the current value of a certain condition can be correlated to its previous value as well as other relevant conditions obtained from the last inspection. Mathematically, we can use the following second-degree polynomial regression to capture the relationship:

$$C_{1,t} = \beta_{0,1} + \beta_{1,1}C_{1,t-1} + \beta_{1,2}C_{2,t-1} + \beta_{1,3}C_{3,t-1} + \beta'_{1,1}C^2_{1,t-1}$$
$$+ \beta'_{1,2}C^2_{2,t-1} + \beta'_{1,3}C^2_{3,t-1} \quad (5)$$

where $C_{1,t-1}$ is the value of the target condition attribute from last inspection; $C_{2,t-1}, C_{3,t-1}\ldots$ are the values of other condition attributes that are related to the target condition attribute. The relevant conditions can be correlated among themselves. Below is an example showing the mathematical relationship of 3 correlated conditions $C_1, C_2$ and $C_3$:

$$\begin{bmatrix} C_{1,t} \\ C_{2,t} \\ C_{3,t} \end{bmatrix} = \begin{bmatrix} \beta_{1,1} & \beta_{1,2} & \beta_{1,3} \\ \beta_{2,1} & \beta_{2,2} & \beta_{2,3} \\ \beta_{3,1} & \beta_{3,2} & \beta_{3,3} \end{bmatrix} \begin{bmatrix} C_{1,t-1} \\ C_{2,t-1} \\ C_{3,t-1} \end{bmatrix}$$
$$+ \begin{bmatrix} \beta'_{1,1} & \beta'_{1,2} & \beta'_{1,3} \\ \beta'_{2,1} & \beta'_{2,2} & \beta'_{2,3} \\ \beta'_{3,1} & \beta'_{3,2} & \beta'_{3,3} \end{bmatrix} \begin{bmatrix} C^2_{1,t-1} \\ C^2_{2,t-1} \\ C^2_{3,t-1} \end{bmatrix} + \begin{bmatrix} \beta_{0,1} \\ \beta_{0,2} \\ \beta_{0,3} \end{bmatrix} \quad (6)$$

### C. Model Combination and Parameter Estimation

One significant advantage of the above discussed models is that they are straightforward and have clear physical meanings compared to "black-box" type of machine learning models. Therefore, the established models are interpretable and can also be confirmed and adjusted by utility asset engineers. This further suggests utility asset engineers can also assume a model based on typical industry values or asset knowledge, without having to rely on actual historical data. In the end, the data-driven condition degradation models, condition correlation models and empirical condition models can all be combined as one model. Mathematically, it is given as:

$$C_{combo} = \sum_{n=1}^{P} \lambda_n C_n^a + \sum_{m=1}^{Q} \lambda_m C_m^c \quad (7)$$

where $P$ and $Q$ are the numbers of condition degradation models and condition correlation models; $\lambda_n$ and $\lambda_m$ are the models' weighting factors; $C_n^a$ and $C_m^c$ are the condition results from individual correlation models.

This is a very powerful concept and provides great flexibility to asset condition generation modelling because:

1. it means the degradation process of a condition attribute can be characterized by different condition degradation models such as a combined linear degradation model and a power degradation, as exemplified in Fig.2. In this example, the degradation shows a linear-like process before 60-year old; but after 60, it becomes more exponential-like;

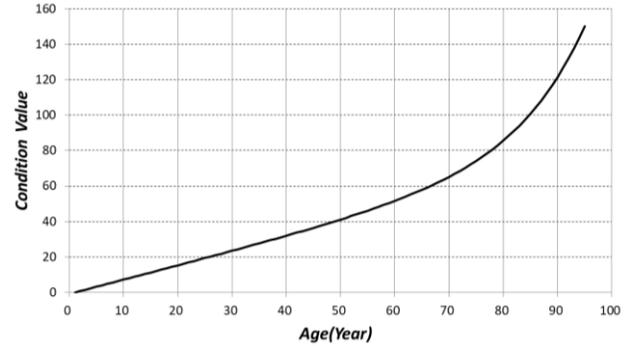

Fig.2. Combined Condition Degradation Model

2. it means the $a$ condition attribute can be inferred from different condition correlation models that consider different sets of possible relevant conditions. This is helpful when users are not certain about the relationships among conditions;
3. it means both the asset age and the last inspection's condition data can be incorporated. However, when only the age data is available, the weighting factor $\lambda_m$ can be just set to 0 so that condition correlation models will not be included;
4. it means empirical models provided by utility asset engineers can also be incorporated. For example, a utility asset engineer may believe there is a specific linear relationship between a contribution attribute and the asset age based on years of observation. This model can be directly included to compensate for the lack of data in certain cases.

To determine the combined model, a two-step parameter estimating process can be used: at the first step, the parameters for individual models can be easily determined through typical least-square based curve-fit methods; then the optimal weighting factors $\lambda_n$ for the linear combination model can also be determined in the same way. It should be noted an asset engineer can choose to subjectively set a model's weighting factor and only allow the rest of weighting factors to be determined mathematically.

*D. Probabilistic Diversification*

The real condition data would not perfectly follow the models established above. Uncertainties that cannot be easily modeled can cause the conditions to vary randomly. Therefore to generate close-to-real data, it is expected that a certain level of variation can be captured to reflect the randomness. Variation can be generated from Gaussian distribution. Each age can have a distinct Gaussian distribution, as illustrated in Fig.3. Mathematically a probabilistic condition value at age $x$ follows:

$$C_x^N \sim G(C_x^R, \sigma_x) \qquad (8)$$

where $G(C_x^R, \sigma_x)$ is the Gaussian noise generator characterized by the expectation $C_x^R$ (i.e. the theoretical value) and standard deviation $\sigma_x$. $\sigma_x$ can be easily determined from historical values. However if a certain age lacks enough data, $\sigma_x$ can also be approximated by adjacent values such as $\sigma_{x-1}$ or $\sigma_{x+1}$. $\sigma_x$ can also be set subjectively as a certain percentage (e.g. 5%) of $C_x^R$ when no historical data is available. This is also applicable when generating hypothetical asset condition data for research purposes. Once the numerical condition becomes probabilistic, a Monte-Carlo Simulation (MSC) can be used to draw the condition value from the Gaussian distribution for a given age.

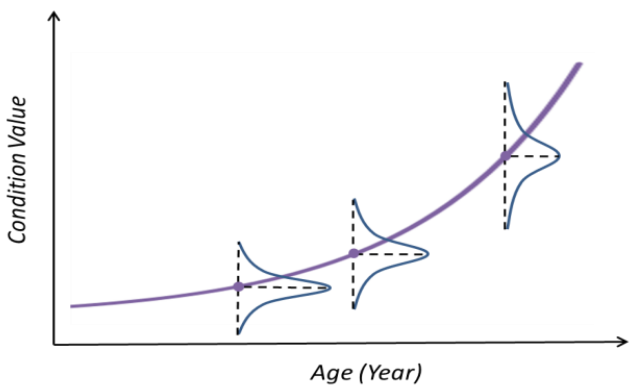

Fig.3. Gaussian distributions for every age for probabilistic diversification

### III. NON-NUMERICAL ASSET CONDITION DEGRADATION MODELS

Non-numerical conditions are in the form of discrete ratings such as low to high and can be converted to numerical conditions by using the equation below [17]:

$$x = \frac{i - 1/2}{N}, i = 1, 2 \ldots N \qquad (9)$$

where $N$ is the total number of ratings and $i$ is the order of the rating. After conversion, non-numerical conditions can be modeled by condition degradation and condition correlation models. However, the results should be converted back to rating orders.

In addition, non-numerical asset conditions can also be directly modeled probabilistically by using the categorical distribution. At each age, the probabilities of the three levels add up to 1.0. Similar to probabilistic diversification for numerical condition data, a categorical distribution can be determined for each age. Mathematically, the model is given as below:

$$\begin{cases} p_i(Age = x) = \dfrac{n_x^i}{n} \\ \sum_{i=1}^{N} p_i = 1 \end{cases} \qquad (10)$$

where $N$ is the total number of ratings; $n_x^i$ is the number of assets that belong to rating level $i$ when age is equal to $x$; $n$ is the total number of existing in-group assets to be analyzed for modelling; $p_i$ is the probability for discrete rating level. It should be noted that similar to numerical asset conditions, empirical discrete probability distributions can also be used to compensate the situation when existing data is not sufficient. If a certain age lacks enough data to produce the categorical probability distribution, average probability distribution of nearby age points can be used.

### IV. VALIDATION

To prove the validity of the above proposed method, two public power asset condition and health index datasets acquired from North American and European utility companies are used for testing [18-19]. Both datasets contain inspection condition records for multiple inspection years. They also include some health index records in which assets with specific conditions have been classified as having certain health indices. Reference [18] is for 45-ft wood power poles and contains inspection records from 1998,2008 and 2018 (i.e. 10-year inspection interval); [19] is for 20 KV XLPE cable segments and contains inspection records from 2010,2013,2016 and 2019 (i.e. 3-year inspection interval). Each dataset contains multiple numerical and non-numerical condition attributes and are ideal for the validation work. The three pole shell-thickness conditions have values declining with age and are converted to values increasing with age to suit the previously discussed condition degradation models. The validation process was carefully designed to test the proposed method from a few different perspectives:

- Test I: Given asset age and last inspection's asset condition data, validate the proposed method can accurately generate the following year's condition data;
- Test II: Given only asset age, validate the proposed method can produce a condition distribution similar to the actual condition distribution;
- Test III: Given asset age and last inspection's asset condition data, validate the proposed method can accurately generate asset health indices that are determined by all condition attributes together;
- Test IV: Compare the performance of the proposed method with different sizes of existing available data.

The 4 tests are discussed in detail as below.

*A. Test I*

For each dataset, 1000 samples are randomly drawn from different years to establish the models (training set) and another 1000 samples are drawn for testing (test set). 3 metrics are used for accuracy evaluation. Mean absolute percent error ($MAPE$) is used to as the error between each actual numerical condition attribute value and the corresponding generated value. Mathematically, they are given as below:

$$MAPE = \frac{1}{m}\sum_{i=1}^{m}\left|\frac{C-\hat{C}}{C}\right| \times 100\% \quad (11)$$

TABLE I: TEST I RESULTS

| Asset Group | Conditions | Condition Attribute Type | Age-driven Condition Models | Conditions in Correlation Models | KL Divergence | Benchmark KL Divergence | MAPE | CMP |
|---|---|---|---|---|---|---|---|---|
| Cable | Partial Discharge(PD) | Numerical | Exponential; Power | PD | 0.011 | 0.193 | 2.75% | N/A |
| | TD Stability(TDS) | Numerical | Linear; Exponential | TDS,DTD,MTD | 0.003 | 0.065 | 3.26% | N/A |
| | Delta TD(DTD) | Numerical | Exponential; Power; | TDS,DTD,MTD | 0.010 | 0.135 | 3.22% | N/A |
| | Mean TD(MTD) | Numerical | Linear; Exponential | TDS,DTD,MTD | 0.009 | 0.043 | 2.77% | N/A |
| | Neutral Corrosion(NC) | Numerical | Logarithmic; Power | NC | 0.022 | 0.042 | 2.77% | N/A |
| | Visual Condition(VC) | Non-Numerical | Linear | N/A | N/A | N/A | N/A | 3.4% |
| Pole | Shell Thickness 1(ST1) | Numerical | Exponential; Power | ST1,ST2,ST3,GC | 0.17 | 0.22 | 6.90% | N/A |
| | Shell Thickness 2(ST2) | Numerical | Exponential; Power; | ST1,ST2,ST3,GC | 0.09 | 0.23 | 7.07% | N/A |
| | Shell Thickness 3(ST3) | Numerical | Exponential; Power | ST1,ST2,ST3,GC | 0.12 | 0.27 | 6.79% | N/A |
| | Ground Circumference(GC) | Numerical | Linear; Exponential | ST1,ST2,ST3,GC | 0.03 | 0.15 | 3.81% | N/A |
| | Surface Condition(SC) | Non-Numerical | Categorical Distribution | N/A | N/A | N/A | N/A | 3.7% |
| | Wood Pecker Hole(WPH) | Non-Numerical | Categorical Distribution | N/A | N/A | N/A | N/A | 2.1% |

For a non-numerical condition attribute, the Condition Mismatch Percentage ($CMP$) is proposed to evaluate the error between the actual and the generated values. The condition ratings are discrete values. Mathematically, $MP$ is defined as the total of rating differences divided by the total number of records:

$$CMP = \frac{1}{m}\sum_{i=1}^{m}|C-\hat{C}| \times 100\% \quad (12)$$

In addition to comparing individual condition values, it is also important to compare the overall distributions of actual condition values and generated condition values. Kullback–Leibler divergence (KL divergence, also known as relative entropy) can be used for this purpose [20]. KL-divergence is a metric to describe how one probability distribution differs from another. KL-divergence has been used to evaluate the similarity of real and generated data in other research works such as Generated Adversarial Network [21]. Mathematically, KL-divergence can be calculated as below:

$$D_{DL}(P||Q) = \sum_{x\in X} P(x) ln\frac{P(x)}{Q(x)} \quad (13)$$

where $P(x)$ is the real condition data's probability distribution; $Q(x)$ is the generated condition data's probability distribution. $D_{DL}(P||Q)$ is greater than 0. When $D_{DL}(P||Q)$ is zero, it means the two probability distributions are exactly the same. Therefore, a smaller $D_{DL}(P||Q)$ indicates that the generated data is able to capture the composition characteristics of the actual data. We applied KL-divergence to all 1000 testing assets which cover the entire age spectrum and hope a small KL-divergence between the actual and generated data can be produced. For benchmark purposes, we also applied KL-divergence to a reference data generator that uniformly draws condition values between the maximum and minimum condition values of a condition attribute, without using any other information.

Based on some preliminary asset analysis, the model specifications for each condition attribute can be quickly set up as listed in Table I: condition degradation models are selected after judging from the patterns of degradation acceleration. For example, PD condition has an accelerated degradation process with age and according to Section II-A, exponential and power models are selected; it is assumed the TDS, DTD,MTD for cables have strong correlation to each other since they are all cable tangent-delta conditions that can be acquired under the same type of cable test; it is also assumed that the 3 remaining pole-shell thicknesses measured from 3 different angles and the ground circumference are correlated since they are all related to wood decay; finally, numerical conditions with no additional condition are only correlated to their own values from their last inspection.

As can be seen, all models are able to generate data with satisfactory accuracy measured by $MAPE$, $R^2$ and $MP$. Also, the KL divergence values are very close to 0 and are also much smaller than the benchmark value calculated from uniformly generating random values between maximum and minimum condition values.

### B. Test II

This test aims to generate condition data only based on asset age, assuming last years' conditions are not available. This means the condition correlation models are not used. Table II summarizes the results by using the same age-driven condition models listed in Table I. As can be seen, individual condition value errors became larger. This is because relying only on age means condition differences from last inspection that are specific to individual assets are not considered. However, KL divergence which focuses on evaluating the overall data distributions remains pretty low, compared to the benchmark KL divergence. This means even by using age, one can still generate useful asset condition data for some occasions. For example, without knowing the individual asset conditions from last inspection, multiple datasets can be generated through Gaussian distributions in the probabilistic diversification step and the average reliability risk can be estimated and taken; hypothetical asset condition data can be quickly generated only based on certain initial age assumptions to support power system and asset reliability research.

TABLE II: TEST II RESULTS

| Asset Class | Conditions | KL Divergence | Benchmark KL Divergence | MAPE | CMP |
|---|---|---|---|---|---|
| Cable | PD | 0.039 | 0.193 | 5.99% | N/A |
| | TDS | 0.051 | 0.065 | 3.59% | N/A |
| | DTD | 0.034 | 0.135 | 4.15% | N/A |
| | MTD | 0.037 | 0.043 | 3.54% | N/A |
| | NC | 0.037 | 0.042 | 3.71% | N/A |
| | VC | N/A | N/A | N/A | 4.7% |
| Pole | ST1 | 0.19 | 0.22 | 7.44% | N/A |
| | ST2 | 0.2 | 0.23 | 7.98% | N/A |
| | ST3 | 0.19 | 0.27 | 6.63% | N/A |
| | GC | 0.15 | 0.15 | 6.15% | N/A |
| | SC | N/A | N/A | N/A | 4.8% |
| | WPH | N/A | N/A | N/A | 3.5% |

*C. Test III*

As shown in Table I, cable and pole each has 6 condition attributes. Test I and Test II focus on evaluating each condition attribute. It is also desired to check the integrated effect of all 6 condition attributes. For example, all condition attributes can be used together to produce a health index that describes overall how healthy the corresponding asset is. In test 3, we evaluate the accuracy of health indices produced from the generated condition data against the actual condition data. Health indices are usually assigned by human experts following certain rules such as the ones described in [15] and the rules can vary from asset to asset and from company to company. For these two datasets, in addition to asset condition records, there are also two sets of health index records. Although no health index rules are directly provided, machine learning can be applied to extract the rules from existing health index records and apply to new assets.

Since the cable dataset uses continuous health index values (0 to 100), an XGBoost regressor is used to extract health index rules [22]; since the pole dataset uses discrete health index levels in 2018 (1 to 5), an XGBoost classifier is used to extract the health index rules. On the basis of Test I, the trained XGBoost classifier and regressor are then applied to both the actual conditions and generated conditions of the 1000 assets in the test sets. As a result, two sets of health indices can be produced and the set based on the actual conditions are used as ground truth for comparison. Similar to $CMP$, a Health Index Mismatch Percentage (HIMP) is defined to compare the discrete health indices for poles. For cables, MAPE is used for continuous value evaluation.

Table III summarizes the comparison results. It shows small HIMP and MAPE which indicate the generated assets conditions can holistically represent the overall statuses of assets. This is an important observation as it implies that the proposed approach can be used for further reliability or health studies, as illustrated in Section V.

TABLE III: ASSET HEALTH INDEX COMPARISON

| Asset Class | HIMP | MAPE |
|---|---|---|
| Cable | N/A | 2.1% |
| Pole | 1.2% | N/A |

*D. Test IV*

Test IV reveals the sensitivity of the proposed method to training data sizes. As discussed in Section I, we hope to be able to generate good condition data with limited amount of existing data. In Test I to III, we have used 1000 training samples to establish the models and in Test IV we vary the data sizes from 50 to 1000. Three conditions are used in the test. As shown in Fig.4, it is found the condition accuracy quickly rises to a high level with only 100 samples and has only marginal or no improvement as the data sizes increases. This means the proposed methods do not require a lot of existing data to achieve satisfactory accuracy. There are three reasons to explain this significant advantage:

1. Common condition degradation models widely used in reliability engineering are adopted in modelling. This is different from the "black-box" type of machine learning methods which have to learn everything from random initialization with no pre-trained models;
2. Human judgement can be incorporated in the modelling process. For example, as discussed in Test 1, the condition degradation models are selected based on observed degradation acceleration patterns; the correlated attributes are found based on some preliminary asset analysis. Also, as discussed in Section II, empirical degradation models can be directly assumed and incorporated. In comparison, machine learning methods cannot easily incorporate this level of pre-knowledge and has to learn everything purely from data;
3. The established models do not have a large number of parameters to tune and are much simpler than widely used machine learning models such as Neural Network and XGBoost. When having many model parameters and small training set, machine learning tend to run into overfitting problems.

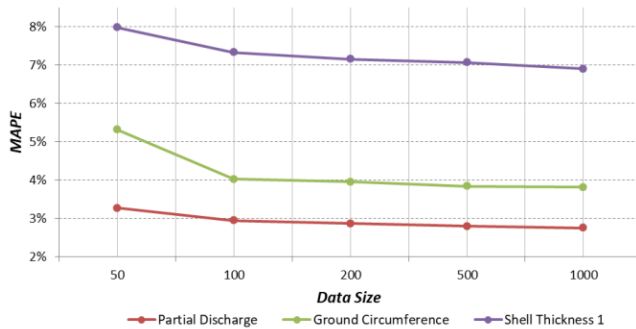

Fig.4. MAPE with increased data sizes

## V. APPLICATION FOR POWER SYSTEM RELIABILITY ASSESSMENT

This section presents a detailed application example to illustrate how the proposed method for generating in-group asset condition data can benefit area reliability assessment work in practice. Generally, the application steps are illustrated in Fig.5.

The IEEE 128-node distribution feeder system was used for illustration [23]. It is assumed all lines between nodes are served by 1200 20kV XLPE underground cable segments with approximately 10m (9.96m) average length according to the total line length of the system.

To create a practical case with actual data, 1200 cable segments are randomly selected from the 20kV XLPE underground cable dataset's year 2019 data [19]. Their health indices can be calculated by using the XGBoost regressor that has been established in Section III.C. In a real utility application, the above step can be understood as getting condition attributes of an asset group in the planning area through proactive inspection. Their HI composition is 31% for level 1, 18% for level 2, 29% for level 3, 20% for level 4 and 2% for level 5.

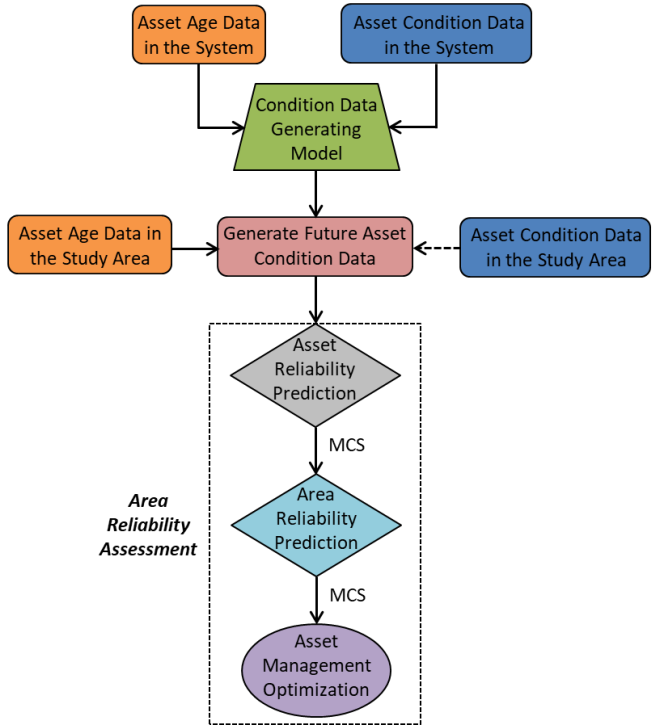

Fig.5. Flowchart of the application of the proposed method to power system reliability assessment

The models established in Section IV.A are used in here to generate future conditions. In a real utility application, these models can be established from available historical data that may come from other planning areas in the system, utility asset engineers' empirical knowledge or a combination of both. The study starting year is 2019.

### A. Cable Health Prediction

Since the 20kV XLPE underground cable was inspected with a 3-year interval, the next three future inspection years should be 2022, 2025 and 2028. First, the condition attribute values in these three inspection years were generated sequentially by using the previously established models; second, the XGBoost regressor was applied to produce the health indices for each asset in each of the inspection years; in the end, for the years between these inspection years, health indices are estimated using linear interpolation. The above steps lead to producing a 10-year health index dataset for the 1200 cable segments, as summarized in Fig.6.

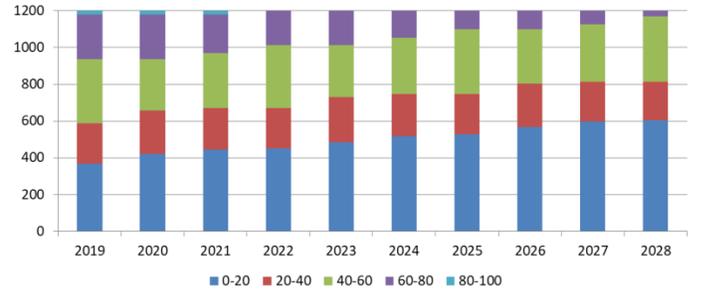

Fig.6. Predicted 10-year health indices for the cable segments

### B. Area Reliability Risk Quantification

Total ownership cost is a commonly used metric to quantify area reliability risk in a monetized way. In the asset management context, it can be mathematically defined as below [24]:

$$TOC = PRC + RRC + FC \qquad (14)$$

here $TOC$ is the total ownership cost; $PRC$ is the proactive replacement cost; $RC$ is the reactive replacement cost; $FC$ is the failure cost.

For example, when a cable segment fails, it will result in an energy loss, which can be converted to a failure cost ($FC$); at the same time, to restore power service, replacing this failed cable segment will result in a reactive cable replacement cost ($RRC$); on the other hand, the utility company can decide to take certain proactive replacement strategies such as proactively replacing unhealthy cable segments before they fail to reduce some potential failure cost at a certain cost ($PRC$). Total ownership cost is the summation of proactive replacement cost (i.e. investment cost), failure cost and reactive replacement cost over a planning period such as 10 years. The best asset management strategy can ensure the total ownership cost is minimal.

In this example, we first assume the utility company decides to take a "run to failure" strategy with no proactive replacement taken. Table IV summarized the assumptions including 5 aging failure probabilities corresponding to 5 health index levels. It is also assumed proactively replacing a cable segment would not result in an economic loss because the utility company can inform and schedule a proper time with the corresponding customer loads to replace the cable with minimum impact.

TABLE IV. ASSUMPTIONS FOR RELIABILITY RISK

| Assumption | Value |
|---|---|
| Annual aging failure probability for HI level 1 cable (0-20) | 10% |
| Annual aging failure probability for HI level 2 cable (20-40) | 5% |
| Annual aging failure probability for HI level 3 cable (40-60) | 2% |
| Annual aging failure probability for HI level 4 cable (60-80) | 1% |
| Annual aging failure probability for HI level 5 cable (80-100) | 0.5% |
| Value of Lost Energy ($/MWH) | 10000 |
| Failure Restoration Duration (hour) | 1 |
| Unit cable replacement cost ($) | 500 |

When a pole fails, a new pole with a HI of 100 will replace it. The yearly based Sequential Monte-Carlo Simulation (SMCS) was adopted to simulate this dynamic process [25]. Value of Lost Energy is used to calculate the failure cost. Fig.7 shows the estimated results including total value of lost energy, reactive replacement cost and total ownership cost over 10 years.

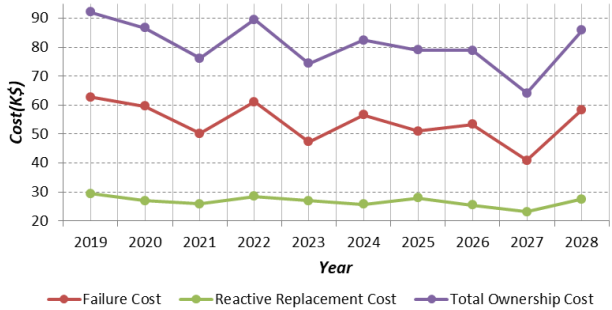

Fig.7. Predicted failure cost, reactive replacement cost and total ownership cost

*C. Asset Replacement Strategy Optimization*

Many types of asset replacement strategies can be taken in order to minimize the total ownership cost. In here, it is assumed that the selected strategy type is to replace X most unhealthy poles (with lowest health indexes) every year and the optimization goal is to find out the optimal X that leads to minimum total ownership cost. Again, SMCS was applied to a number of X scenarios and Fig.8 shows the estimation results. As can be seen, the utility company should proactively replace 70 poles every year to achieve an estimated minimal total ownership cost of $ 702,580 from 2019 to 2028 for this feeder system.

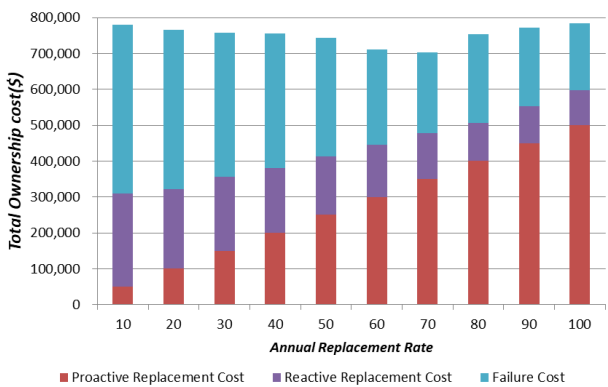

Fig.8. Total ownership cost and cost compositions vs. annual replacement rate

## VI. DISCUSSIONS AND CONCLUSIONS

This paper presents a novel method to generate in-group asset condition data for power system reliability assessment. The method uses condition degradation, condition correlation, categorical distribution models and their combinations with probabilistic diversification to generate realistic numerical and non-numerical in-group asset condition values. The proposed method was systematically validated by two public datasets with multiple metrics. Also, a detailed reliability assessment example is presented to demonstrate the usefulness of the method step by step.

The models in the proposed method are highly interpretable for human experts and flexible to configure. The performance does not rely on a large volume of existing data. The main significances of the proposed method are:

- Current and future in-group asset conditions can be generated based on limited historical data and/or human expert knowledge. Such data can effectively support asset reliability assessment work such as asset and system reliability risk quantification and asset management optimization;
- It allows researchers to generate hypothetical asset condition data conveniently in the field of power system and asset reliability research for various research purposes.

**Ming Dong** (SM'18) received his Ph.D degree from Department of Electrical and Computer Engineering, University of Alberta and the Certificate of Data Science and Big Data Analytics from Massachusetts Institute of Technology. To date, he has been working at two major utility companies and an ISO in Canada as a Senior Engineer and Senior Data Scientist for 8 years. He is an editor of IEEE TRANSACTIONS ON POWER DELIVERY. He is the chair of 2020 IEEE PES General Meeting's panel session on "Advanced Analytics for Power Asset Management". His research has been focused on the applications of artificial intelligence and big data analytics to power system asset management, planning and operations.

**Alexandre Nassif** (SM'13) is a specialist engineer responsible for the Asset Management Distribution capital investment program at LUMA Energy. He is also ATCO's SME in power quality, micro-grid planning, and distributed energy resource integration. Before joining these organizations, he worked for Hydro One (Toronto) as a protection planning engineer, and simultaneously for Ryerson University as a post-doctoral research fellow. He has published more than 80 technical papers in international journals and conferences and is an editor of IEEE TRANSACTIONS ON POWER DELIVERY. He is a professional engineer in the provinces of Alberta and Nova Scotia. He holds a PhD degree from the University of Alberta.

**Wenyuan Li** (F'02-LF'18) is currently a professor with Chongqing University, Chongqing, China. His research interests include power system asset management, planning, operation, optimization, and reliability assessment. He is a Fellow of the Canadian Academy of Engineering and the Engineering Institute of Canada, and a Foreign Member of the Chinese Academy of Engineering. He was a recipient of several IEEE PES awards, including the IEEE PES Roy Billinton Power System Reliability Award, in 2011, and the IEEE Canada Electric Power Medal, in 2014.